\documentclass[11pt,twoside]{article}
\usepackage{asp2004}
\usepackage{psfig}
\reversemarginpar

\begin{document}
\title{Young Massive Clusters in Merging and Starburst Galaxies}

\author{Richard de Grijs}
\affil{Department of Physics \& Astronomy, University of Sheffield,
Hicks Building, Hounsfield Road, Sheffield S3 7RH, UK}

\begin{abstract}
The currently available empirical evidence on the star formation processes in
the extreme, high-pressure environments induced by galaxy encounters, mostly
based on high-resolution {\sl Hubble Space Telescope} imaging observations,
strongly suggests that star {\it cluster} formation is an important and
perhaps even the dominant mode of star formation in the starburst events
associated with galaxy interactions. The production of young massive star
clusters (YSCs) seems to be a hallmark of intense star formation, particularly
in interacting and starburst galaxies. Their sizes, luminosities, and mass
estimates are entirely consistent with what is expected for young Milky
Way-type globular clusters (GCs). YSCs are important because of what they can
tell us about GC formation and evolution (e.g., initial characteristics and
early survival rates). They are also of prime importance as probes of the
formation and (chemical) evolution of their host galaxies, and of the initial
mass function in the extreme environments required for cluster formation.
Recent evidence lends support to the scenario that Milky Way-type GCs
(although more metal rich), which were once thought to be the oldest building
blocks of galaxies, are still forming today.
\end{abstract}

\section{Extreme environmental conditions}

Stars rarely form in isolation. In fact, it is well known that the
vast majority of stars in the Galaxy, and also in nearby galaxies, are
found in groups ranging from small associations, containing some 100
M$_\odot$, to compact, old ``globular'' and young massive
clusters. The nearest examples of these latter objects include the
Galactic star-forming regions NGC 3603 and Westerlund 1, and the giant
starburst region 30 Doradus with its central star cluster R136 in the
Large Magellanic Cloud.

Although the older Galactic {\it open} clusters (with ages of several
Gyr) are undoubtedly gravitationally bound objects, their lower masses
compared to the globular cluster population, and more diffuse
structures make them much more vulnerable to disk (and bulge) shocking
when they pass through the Galactic disk (or close to the bulge), thus
leading to enhanced cluster evaporation. These objects are therefore
unlikely globular cluster progenitors. It appears that the conditions
for the formation of compact, massive star clusters -- that have the
potential to eventually evolve into globular cluster-type objects by
the time they reach a similar age -- are currently not present in the
Galaxy, or at best to a very limited extent (e.g., Westerlund 1;
Hanson 2003).

The production of luminous, massive yet compact star clusters seems to
be a hallmark of the most intense star-forming episodes, or
starbursts. The defining properties of young massive star clusters
(YSCs; with masses often significantly in excess of $M_{\rm cl} = 10^5
{\rm M}_\odot$) have been explored in intense starburst regions in
several dozen galaxies, often involved in gravitational interactions
of some sort (e.g., Holtzman et al. 1992, Whitmore et al. 1993,
O'Connell et al. 1994, Conti et al. 1996, Watson et al. 1996, Carlson
et al. 1998, de Grijs et al.  2001, 2003a,b,c,d,e).

An increasingly large body of observational evidence suggests that a
large fraction of the star formation in starbursts actually takes
place in the form of such concentrated clusters, rather than in
small-scale star-forming ``pockets''. YSCs are therefore important as
benchmarks of cluster formation and evolution. They are also important
as tracers of the history of star formation of their host galaxies,
their chemical evolution, the initial mass function (IMF), and other
physical characteristics in starbursts.

In a detailed study of the young star cluster population associated
with the fading starburst region ``B'' in the prototype nearby
starburst galaxy M82 (de Grijs et al. 2001), we concluded that the
last tidal encounter between M82 and its large neighbour spiral galaxy
M81, which occurred about $500 - 800$ Myr ago (Brouillet et al. 1991,
Yun 1999) had a major impact on what was probably an otherwise normal,
quiescent, disk galaxy. It caused a concentrated starburst, resulting
in a pronounced peak in the clusters' age distribution, roughly 1 Gyr
ago (de Grijs et al. 2001, 2003c, Parmentier et al.  2003). The
enhanced cluster formation decreased rapidly, within a few hundred Myr
of its peak. However, general star formation activity continued in the
galactic disk of M82's region B, probably at a much lower rate, until
$\sim 20$ Myr ago.

The evidence for the decoupling between cluster and field star
formation is consistent with the view that star cluster formation
requires special conditions, such as large-scale gas flows, in
addition to the presence of dense gas (cf.  Ashman \& Zepf 1992,
Elmegreen \& Efremov 1997).  Such conditions occur naturally in the
extreme environments of gravitationally interacting galaxies.

Using optical observations of the ``Mice'' and ``Tadpole'' interacting
galaxies (NGC 4676 and UGC 10214, respectively) -- based on a subset
of the Early Release Observations obtained with the {\sl Advanced
Camera for Surveys} on board the {\sl Hubble Space Telescope (HST)} --
and the novel technique of pixel-by-pixel analysis of their
colour-colour and colour-magnitude diagrammes, we deduced the systems'
star and star cluster formation histories (de Grijs et al. 2003e).

In both of these interacting systems we find several dozen YSCs (or,
alternatively, compact star-forming regions), which overlap spatially
with regions of active star formation in the galaxies' tidal tails and
spiral arms (from a comparison with H$\alpha$ observations that trace
active star formation; Hibbard \& van Gorkom 1996). We estimate that
the main gravitational interactions responsible for the formation of
these clusters occurred $\sim 150 - 200$ Myr ago.

In fact, we show that star cluster formation is a major mode of star
formation in galactic interactions, with $\ge 35$\% of the active star
formation in encounters occurring in star clusters (de Grijs et
al. 2003e). The tidal tail of the Tadpole system is dominated by star
forming regions, which contribute $\sim 70$\% of the total flux in the
{\sl HST} {\it I}-band filter (decreasing to $\sim 40$\% in the
equivalent {\it B}-band filter). If the encounter occurs between
unevenly matched, gas-rich galaxies then, as expected, the effects of
the gravitational interaction are much more pronounced in the smaller
galaxy. For instance, when we compare the impact of the interaction as
evidenced by star cluster formation between M82 (de Grijs et al. 2001,
2003b,c) and M81 (Chandar et al.  2001), or the star cluster formation
history in the ``Whirlpool Galaxy'' M51 (Bik et al. 2003), which is
currently in the process of merging with the smaller spiral galaxy NGC
5194, the evidence for enhanced cluster formation in the larger galaxy
is minimal if at all detectable.

The NGC 6745 system represents a remarkably violently star-forming
interacting pair of unevenly matched galaxies.  The optical morphology
of NGC 6745, and in particular the locations of the numerous bright
blue star-forming complexes and compact cluster candidates, suggest a
recent tidal passage by the small northern companion galaxy (NGC
6745c; nomenclature from Karachentsev et al.  1978) across the eastern
edge of the main galaxy, NGC 6745a.  The high relative velocities of
the two colliding galaxies likely caused ram pressure at the
surface of contact between both galaxies, which -- in turn -- is
responsible for the triggering of enhanced star and cluster formation,
most notably in the interaction zone in between the two galaxies, NGC
6745b (cf.  de Grijs et al.  2003a).  The smaller galaxy, however,
does not show any significant enhanced cluster formation, which is
most likely an indication that it contains very little gas.

For the NGC 6745 young cluster system we derive a median age of $\sim
10$ Myr.  Based on the age distribution of the star clusters, and on
the H{\sc i} morphology of the interacting system, we confirm the
interaction-induced enhanced cluster formation scenario once
again. NGC 6745 contains a significant population of high-mass
clusters, with masses in the range $6.5 \le \log( M_{\rm cl}/{\rm
M}_\odot ) \le 8.0$. These clusters do not have counterparts among the
Galactic globular clusters (e.g., Mandushev et al. 1991, Pryor \&
Meylan 1993), but are similar to or exceed the spectroscopically
confirmed mass estimates of the ``super star clusters'' (SSCs) in M82
(e.g., M82 F and L; see Smith \& Gallagher 2001) and the Antennae
galaxies (Mengel et al.  2002). We caution, however, that these
massive SSC candidates may not be gravitationally bound objects, but
more diffuse star forming regions or aggregates of multiple unresolved
clusters instead. Nevertheless, we measure a very compact effective
radius for the most massive object ($M_{\rm cl} \simeq 5.9 \times 10^8
{\rm M}_\odot$) of only $R_{\rm eff} \sim 16$ pc. However, this object
appears very elongated, or may in fact be a double cluster. We should
also keep in mind that this high mass estimate is a strong function of
the (low) metallicity assumed; if we had assumed (higher) solar
metallicity for this object, the derived age would have been
significantly smaller ($\sim 10-20$ Myr vs. $\sim 1$ Gyr assumed in
our study), and the mass could be smaller by a factor of $\ga
10$. Even so, if we could confirm this mass estimate
spectroscopically, either of the subcomponents would be the most
massive cluster known to date, significantly exceeding cluster W3 in
NGC 7252, which has a mass of about $(3-18) \times 10^7 {\rm
M}_\odot$, depending on the age, metallicity and IMF assumed
(Schweizer \& Seitzer 1998; Maraston et al.  2001, 2004). Our
detection of such massive SSCs in NGC 6745, which are mostly located
in the intense interaction zone, supports the scenario that such
objects form preferentially in the extreme environments of interacting
galaxies.

\section{An evolutionary connection?} 

The (statistical) derivation of galaxy formation and evolution
scenarios using their star cluster systems as tracers is limited to
the study of integrated cluster properties for all but the nearest
galaxies, even at {\sl HST} spatial resolution.

The question remains, therefore, whether or not at least a fraction of
the YSCs seen in abundance in extragalactic starbursts, are
potentially the progenitors of globular cluster-type objects in their
host galaxies. If we could settle this issue convincingly, one way or
the other, the implications of such a result would have profound and
far-reaching implications for a wide range of astrophysical questions,
including (but not limited to) our understanding of the process of
galaxy formation and assembly, and the process and conditions required
for star (cluster) formation. Because of the lack of a statistically
significant sample of similar nearby objects, however, we need to
resort to either statistical arguments or to the painstaking approach
of one-by-one studies of individual objects in more distant
galaxies. With the ever increasing number of large-aperture
ground-based telescopes equipped with state-of-the-art high-resolution
spectroscopic detectors and the wealth of observational data provided
by the {\sl HST}, we may now be getting close to resolving this
important issue. It is of paramount importance, however, that
theoretical developements go hand in hand with observational advances.

The present state-of-the-art teaches us that the sizes, luminosities,
and -- in several cases -- spectroscopic mass estimates of most
(young) extragalactic star cluster systems are fully consistent with
the expected properties of young Milky Way-type globular cluster
progenitors (e.g., Meurer 1995, van den Bergh 1995, Ho \& Filippenko
1996a,b, Schweizer \& Seitzer 1998, de Grijs et al. 2001, 2003d). For
instance, for the young massive star cluster system in the centre of
the nearby starburst spiral galaxy NGC 3310, we find a median mass of
$\langle \log(M_{\rm cl}/{\rm M}_\odot) \rangle = 5.24 \pm 0.05$ (de
Grijs et al. 2003d); their mass distribution is characterised by a
Gaussian width of $\sigma_{\rm Gauss} \simeq 0.33$ dex. In view of the
uncertainties introduced by the poorly known lower-mass slope of the
stellar IMF ($m_\ast \le 0.5 {\rm M}_\odot$), our median mass estimate
of the NGC 3310 cluster system -- which was most likely formed in a
(possibly extended) global burst of cluster formation $\sim 3 \times
10^7$ yr ago -- is remarkably close to that of the Galactic globular
cluster system (cf. de Grijs et al.  2003d; Fig. 1).

\begin{figure}[h!]
\hspace*{1.5cm}
\psfig{figure=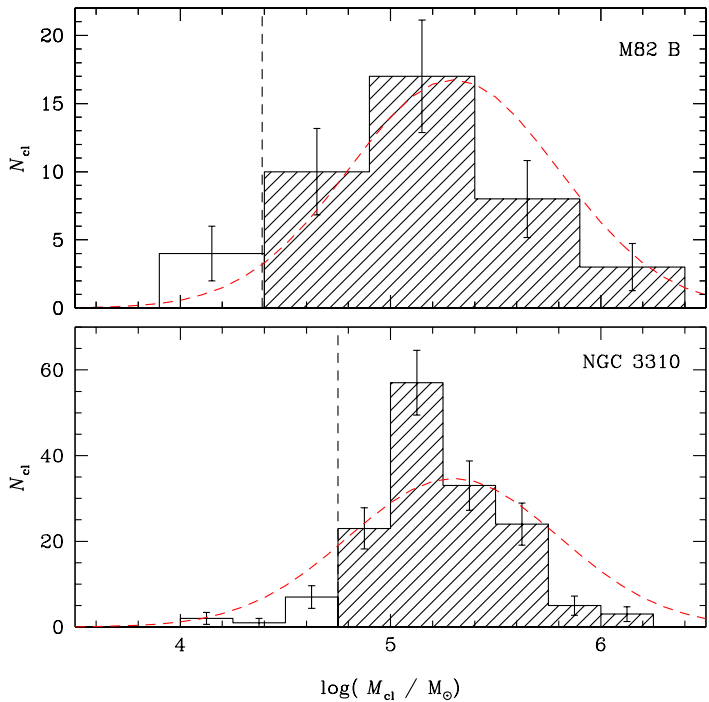,width=10cm}
\caption[]{Mass distributions of the intermediate-age, $\sim 1$
Gyr-old star clusters in M82 B {\it (top)}, and the young ($\sim 30$
Myr-old) cluster population dominating the central regions of the
starburst galaxy NGC 3310 {\it (bottom)}. For comparison, the
dashed Gaussian distributions show the mass distribution of the
Galactic globular cluster system, normalised to the same number of
clusters in each population. The vertical dashed lines indicate our 50
per cent completeness limits, with severe incompleteness affecting the
clusters populating the open histograms.}
\end{figure}

However, the postulated evolutionary connection between the recently
formed massive star clusters in regions of violent star formation and
starburst galaxies, and old globular clusters in the nearby Universe
is still a contentious issue. The evolution and survivability of young
clusters depend crucially on the stellar IMF of their constituent
stars (cf. Smith \& Gallagher 2001): if the IMF is too shallow, i.e.,
if the clusters are significantly depleted in low-mass stars compared
to (for instance) the solar neighbourhood, they will disperse within a
few orbital periods around their host galaxy's centre, and likely
within about a Gyr of their formation (e.g., Smith \& Gallagher 2001,
Mengel et al. 2002).

Ideally, one would need to obtain (i) high-resolution spectroscopy
(e.g., with 8m-class ground-based telescopes) of all clusters in a
given cluster sample in order to obtain dynamical mass estimates
(assuming that the clusters are in full virial equilibrium), and (ii)
high-resolution imaging (e.g., with the {\sl HST}) to measure their
luminosities. One could then estimate the mass-to-light (M/L) ratios
for each cluster, and their ages from the spectra. The final, crucial
analysis would involve a direct comparison between the clusters'
locations in the M/L ratio vs. age diagramme with models of ``simple
stellar populations'' governed by a variety of IMF descriptions
(cf. Smith \& Gallagher 2001, Mengel et al. 2002).

However, individual young star cluster spectroscopy, feasible today
with 8m-class telescopes for the nearest systems, is very
time-consuming, since observations of large numbers of clusters are
required to obtain statistically significant results. Instead, one of
the most important and most widely used diagnostics, both to infer the
star (cluster) formation history of a given galaxy, and to constrain
scenarios for its expected future evolution, is the distribution of
cluster luminosities, or -- alternatively -- their associated masses,
commonly referred to as the cluster luminosity and mass functions
(CLF, CMF), respectively.

Starting with the seminal work by Elson \& Fall (1985) on the young
cluster system in the Large Magellanic Cloud (with ages $\le 2 \times
10^9$ yr), an ever increasing body of evidence seems to imply that the
CLF of YSCs is well described by a power law down to the lowest
luminosities (and thus masses). On the other hand, for the old
globular cluster systems in the local Universe, with ages $\ge 10$
Gyr, the CLF shape is well established to be roughly Gaussian
(Whitmore et al. 1995, Harris 1996, 2001, Harris et al. 1998). This
shape is almost universal, showing only a weak dependence on the
metallicity and mass of the host galaxy (e.g., Harris 1996, 2001,
Whitmore et al. 2002).

This type of observational evidence has led to the popular -- but thus
far mostly speculative -- theoretical prediction that not only a
power-law, but {\it any} initial CLF (and CMF) will be rapidly
transformed into a Gaussian distribution because of (i) stellar
evolutionary fading of the lowest-luminosity (and therefore
lowest-mass) clusters to below the detection limit; and (ii)
disruption of the low-mass clusters due both to interactions with the
gravitational field of the host galaxy, and to cluster-internal
two-body relaxation effects (such as caused by star-star collisions
and the resulting redistribution of mass inside the cluster) leading
to enhanced cluster evaporation (e.g., Elmegreen \& Efremov 1997,
Gnedin \& Ostriker 1997, Ostriker \& Gnedin 1997, Fall \& Zhang 2001).

We recently reported the first discovery of an approximately Gaussian
CLF (and CMF) for the star clusters in M82 B formed roughly
simultaneously in a pronounced burst of cluster formation (de Grijs et
al. 2003b,c). This provides the very first sufficiently deep CLF (and
CMF) for a star cluster population at intermediate age (of $\sim 1$
billion years), which thus serves as an important benchmark for
theories of the evolution of star cluster systems (but see
Goudfrooij, this volume).

The shape of the CLF (CMF) of young cluster systems has recently
attracted renewed theoretical and observational attention. Various
authors have pointed out that for young star clusters exhibiting an
age range, one must first correct their CLF to a common age before a
realistic assessment of their evolution can be achieved (e.g., Meurer
1995, Fritze--v. Alvensleben 1998, 1999, de Grijs et al. 2001,
2003b,c). This is particularly important for young cluster systems
exhibiting an age spread that is a significant fraction of the
system's median age, because of the rapid evolution of the colours and
luminosities of star clusters at young ages (below $\sim 1$
Gyr). Whether the observed power laws of the CLF and CMF for young
star cluster systems are intrinsic to the cluster population or
artefacts due to the presence of an age spread in the cluster
population -- which might mask a differently shaped underlying
distribution -- is therefore a matter of debate (see, e.g., Carlson et
al. 1998, Zhang \& Fall 1999, Vesperini 2000, 2001).

Nevertheless, the CLF shape and characteristic luminosity of the M82 B
cluster system is nearly identical to that of the apparently universal
CLFs of the old globular cluster systems in the local Universe (e.g.,
Whitmore et al. 1995, Harris 1996, 2001, Ashman \& Zepf 1998, Harris
et al. 1998). This is likely to remain virtually unchanged for a
Hubble time, if the currently most popular cluster disruption models
hold. With the very short characteristic cluster disruption time-scale
governing M82 B (de Grijs et al. 2003c), its cluster mass distribution
will evolve towards a higher characteristic mass scale than that of
the Galactic globular cluster system by the time it reaches a similar
age. Thus, this evidence, combined with the similar cluster sizes (de
Grijs et al. 2001), lends strong support to a scenario in which the
current M82 B cluster population will eventually evolve into a
significantly depleted old Milky Way-type globular cluster system
dominated by a small number of high-mass clusters. However, they will
likely be more metal-rich than the present-day old globular cluster
systems.

The connection between young or intermediate-age star cluster systems,
as in M82 B, and old globular clusters lends support to the
hierarchical galaxy formation scenario. Old globular clusters were
once thought to have been formed at the time of, or before, galaxy
formation, i.e., during the first galaxy mergers and
collisions. However, we have now shown that the evolved CLF of the
compact star clusters in M82 B most likely to survive for a Hubble
time will probably resemble the high-mass wing of the ``universal''
old globular cluster systems in the local Universe. Proto-globular
cluster formation thus appears to be continuing until the present.

\section{The bottom line}

In summary, in this review I have shown that young, massive star
clusters are the most significant end products of violent star-forming
episodes (starbursts) triggered by galaxy collisions and gravitational
interactions in general. Their contribution to the total luminosity
induced by such extreme conditions dominates, by far, the overall
energy output due to the gravitationally-induced star formation. The
general characteristics of these newly-formed clusters (such as their
ages, masses, luminosities, and sizes) suggest that at least a
fraction may eventually evolve into equal, or perhaps slightly more
massive, counterparts of the abundant old globular cluster systems in
the local Universe, although they will likely be more metal rich than
the present generation of globular clusters. Establishing whether or
not such an evolutionary connection exists requires our detailed
knowledge of not only the physics underlying the evolution of
``simple'' stellar populations (i.e., idealised model clusters), but
also that of cluster disruption in the time-dependent gravitational
potentials of interacting galaxies. Initial results seem to indicate
that proto-globular clusters do indeed continue to form today, which
would support hierarchical galaxy formation scenarios.  Settling this
issue conclusively will have far-reaching consequences for our
understanding of the process of galaxy formation and assembly, and of
star formation itself, both of which processes are as yet poorly
understood.

\acknowledgments{This work would not have been possible without
valuable contributions by many collaborators, including Uta
Fritze--v. Alvensleben, Peter Anders, Henny Lamers, Nate Bastian, and
Jay Gallagher, to whom I am indebted.}


\begin{thebibliography}{}

\bibitem[]{} Ashman K.M., Zepf S.E., 1992, ApJ, 384, 50
\bibitem[]{} Ashman K.M., Zepf S.E., 1998, Globular Cluster Systems,
Cambridge University Press
\bibitem[]{} Bik A., Lamers H.J.G.L.M., Bastian N., Panagia N., 
Romaniello M., 2003, A\&A, 397, 473
\bibitem[]{} Brouillet N., Baudry A., Combes F., Kaufman M., Bash F.,
1991, A\&A, 242, 35
\bibitem[]{} Carlson M.N., et al., 1998, AJ, 115, 1778
\bibitem[]{} Chandar R., Ford H.C., Tsvetanov Z., 2001, AJ, 122, 1330
\bibitem[]{} Conti P.S., Leitherer C., Vacca W.D., 1996, ApJ, 461, L87
\bibitem[]{} de Grijs R., O'Connell R.W., Gallagher J.S., 2001, AJ, 121,
768
\bibitem[]{} de Grijs R., Anders P., Lynds R., Bastian N., Lamers
H.J.G.L.M., Fritze--v.  Alvensleben U., 2003a, MNRAS, 343, 1285
\bibitem[]{} de Grijs R., Bastian N., Lamers, H.J.G.L.M., 2003b, ApJ,
583, L17
\bibitem[]{} de Grijs R., Bastian N., Lamers H.J.G.L.M., 2003c, MNRAS,
340, 197
\bibitem[]{} de Grijs R., Fritze--v.  Alvensleben U., Anders P.,
Gallagher J.S., Bastian N., Taylor V.A., Windhorst R.A., 2003d, MNRAS,
342, 259
\bibitem[]{} de Grijs R., Lee J.T., Mora Herrera M.C., Fritze--v. 
Alvensleben U., Anders P., 2003e, New Astron., 8, 155
\bibitem[]{} Elmegreen B.G., Efremov Y.N., 1997, ApJ, 480, 235
\bibitem[]{} Elson R.A.W., Fall S.M., 1985, PASP, 97, 692
\bibitem[]{} Fall S.M., Zhang Q., 2001, ApJ, 561, 751
\bibitem[]{} Fritze--v.  Alvensleben U., 1998, A\&A, 336, 83
\bibitem[]{} Fritze--v.  Alvensleben U., 1999, A\&A, 342, L25
\bibitem[]{} Gnedin O.Y., Ostriker J.P., 1997, ApJ, 474, 223  
\bibitem[]{} Hanson M.M., 2003, ApJ, 597, 957
\bibitem[]{} Harris W.E., 1996, AJ, 112, 1487
\bibitem[]{} Harris W.E., 2001, in: Star Clusters, Saas-Fee Advanced
Course 28, Spinger-Verlag, 223
\bibitem[]{} Harris W.E., Harris G.L.H., McLaughlin D.E., 1998, AJ, 115,
1801
\bibitem[]{} Hibbard J.E., van Gorkom J.H., 1996, AJ, 111, 655
\bibitem[]{} Ho L.C., Filippenko A.V., 1996a, ApJ, 466, L83
\bibitem[]{} Ho L.C., Filippenko A.V., 1996b, ApJ, 472, 600
\bibitem[]{} Holtzman J.A., et al., 1992, AJ, 103, 691
\bibitem[]{} Karachentsev I.D., Karachentseva V.E., Shcherbanovskii
A.L., 1978, PAZh, 4, 483 (English transl. in Soviet Astr. Lett., 4,  
261)
\bibitem[]{} Mandushev G., Spassova N., Staneva A., 1991, A\&A, 252,
94
\bibitem[]{} Maraston C., Kissler-Patig M., Brodie J.P., Barmby P., Huchra
J.P., 2001, A\&A, 370, 176
\bibitem[]{} Maraston C., Bastian N., Saglia R.P., Kissler-Patig M.,
Schweizer F., Goudfrooij P., 2004, A\&A, in press (astro-ph/0311232)
\bibitem[]{} Mengel S., Lehnert M.D., Thatte N., Genzel R., 2002, A\&A,
383, 137
\bibitem[]{} Meurer G.R., 1995, Nat., 375, 742
\bibitem[]{} O'Connell R.W., Gallagher J.S., Hunter D.A., 1994, ApJ,
433, 65
\bibitem[]{} Ostriker J.P., Gnedin O.Y., 1997, ApJ, 487, 667
\bibitem[]{} Parmentier G., de Grijs R., Gilmore G.F., 2003, MNRAS, 342,
208
\bibitem[]{} Pryor C., Meylan G., 1993, in: Structure and Dynamics of
Globular Clusters, Djorgovski S.G., Meylan G., eds., San Francisco: ASP,
p.  357
\bibitem[]{} Schweizer F., Seitzer P., 1998, AJ, 116, 2206
\bibitem[]{} Smith L.J., Gallagher J.S., 2001, MNRAS, 326, 1027
\bibitem[]{} van den Bergh S., 1995, Nat., 374, 215
\bibitem[]{} Vesperini E., 2000, MNRAS, 318, 841
\bibitem[]{} Vesperini E., 2001, MNRAS, 322, 247
\bibitem[]{} Watson A.M., et al., 1996, AJ, 112, 534
\bibitem[]{} Whitmore B.C., Schweizer F., Kundu A., Miller B.W., 2002,
AJ, 124, 147
\bibitem[]{} Whitmore B.C., Schweizer F., Leitherer C., Borne K., Robert
C., 1993, AJ, 106, 1354
\bibitem[]{} Whitmore B.C., Sparks W.B., Lucas R.A., Macchetto F.D.,
Biretta J.A., 1995, ApJ, 454, L73
\bibitem[]{} Yun M.S., 1999, IAU Symp. 186: Galaxy Interactions at Low
and High Redshift, Barnes J.E., Sanders D.B., eds., p.  81
\bibitem[]{} Zhang Q., Fall S.M., 1999, ApJ, 527, L81

\end{thebibliography}
\end{document}